\documentclass{aa}  

\usepackage{graphicx}
\usepackage{txfonts}
\usepackage[section]{placeins}

\usepackage{ulem}
\usepackage[usenames,dvipsnames,svgnames,table]{xcolor}

\begin{document} 

\title{Zero-phase angle asteroid taxonomy classification using unsupervised machine learning algorithms}


   \author{M. Colazo
          \inst{1}
          \and
          A. Alvarez-Candal \inst{2,3,4}
          \and
          R. Duffard \inst{2}
          }

   \institute{Instituto de Astronomía Teórica y Experimental, CONICET-UNC, Laprida 854, Córdoba, Argentina\\
              \email{milagros.colazo@mi.unc.edu.ar}\\
         \and
            Instituto de Astrofísica de Andalucía, CSIC, Glorieta de la Astronomía s/n, 18008 Granada, Spain\\
        \and
             Instituto de Física Aplicada a las Ciencias y las Tecnologías, Universidad de Alicante, San Vicent del Raspeig, E03080, Alicante, Spain\\
        \and 
            Observatório Nacional/MCTIC, Rua General José Cristino 77, Río de Janeiro, RJ, 20921-400, Brazil
           }

   \date{Received September 15, 1996; accepted March 16, 1997}

 
  \abstract
   {We are in an era of large catalogs and, thus, statistical analysis tools for large data sets, such as machine learning, play a fundamental role. One example of such a survey is the Sloan Moving Object Catalog (MOC), which lists the astrometric and photometric information of all moving objects captured by the Sloan field of view. One great advantage of this telescope is represented by its set of five filters, allowing for taxonomic analysis of asteroids by studying their colors. However, until now, the color variation produced by the change of phase angle of the object has not been taken into account.}
   {In this paper, we address this issue by using absolute magnitudes for classification. We aim to produce a new taxonomic classification of asteroids based on their magnitudes that is unaffected by variations caused by the change in phase angle.}
   {We selected 9481 asteroids with absolute magnitudes of $H_g$, $H_i$ and $H_z$, computed from the Sloan Moving Objects Catalog using the HG$^*_{12}$ system. We calculated the absolute colors with them. To perform the taxonomic classification, we applied a unsupervised machine learning algorithm known as fuzzy C-means. This is a useful soft clustering tool for working with {data sets where the different groups are not completely separated and there are regions of overlap between them. We have chosen} to work with the four main taxonomic complexes, C, S, X, and V, as they comprise most of the known spectral characteristics.}
   {We classified a total of 6329 asteroids with more than 60$\%$ probability of belonging to the assigned taxonomic class, with 162 of these objects having been characterized by an ambiguous classification in the past. By analyzing the sample obtained in the plane Semimajor axis versus inclination, we identified 15 new V-type asteroid candidates outside the Vesta family region.}
   {}


   \maketitle
%
\section{Introduction}
Astronomical surveys present a challenge every day in terms of generating vast amounts of data that increase exponentially. Nowadays, we can observe the sky from the Earth or space at different wavelengths, targeting different footprints or objects. Moreover, the science of the Solar System is now subject to a unique scenario with the advent of surveys dedicated explicitly to the observation of asteroids, such as the Lincoln Near-Earth Asteroid Research (LINEAR) \citep{St00}, the Catalina Sky Survey (CSS) \citep{Ch14, Ch19}, the NEOWISE \citep{Ma14}, and many more. However, a great deal of data (and discoveries) come from other surveys, whose main objective is not the study of minor bodies: Gaia \citep{Gaia}, Sloan Digital Sky Survey (SDSS) \citep{Yo00}, TESS (Transiting Exoplanet Survey Satellite) \citep{Ri15}, and others. Although the ultimate purpose of these telescopes is not the exploration of asteroids, these objects inevitably cross the field of view of the camera, producing a wealth of data worth exploring.

An example is the Sloan Moving Object Catalog (MOC), a list of astrometric and photometric information on all moving objects detected in Sloan images. Numerous works have made use of this database, for instance: \cite{Ro08, Gi08, As08, So10, Of12, Al13}. It is essential to note that the multi-wavelength nature of the Sloan data allows for detailed studies of a range of parameters. In particular, it allows for studies of the brightness variation of the objects with phase angle, the so-called phase curves. Phase curves provide valuable information, mainly the absolute magnitude of the object, directly related to its size and geometric albedo -- parameters that are essential for dynamical studies on the formation or evolution of the Solar System.

We can exploit the phase curves in full using photometric models to fit the data. Among the many systems available in the literature, one of the better known is the HG system \citep{Bo89}, where $H$ represents the absolute magnitude of the object and $G$ its slope parameter, which describes the shape of the phase function. \cite{Mu10} presented a new photometric system (HG$_{1}$G$_{2}$), now with three parameters, which gives more accurate results if we are dealing with a densely sampled phase curve even for small phase angles \citep{Co21}. In the case of phase curves that are sparsely covered or data with significant uncertainties, \cite{Pe16} proposed the HG$^*_{12}$ system. In a nutshell, the HG$^*_{12}$ system parametrizes \cite{Mu10}'s $G_1$ and $G_2$ such as: $$G_i=a_i+b_iG^*_{12},$$ where $a_i$ and $b_i$ are known constants (see \citealt{Pe16}).

Another type of study that can be carried out thanks to the five-filter observations offered by Sloan is the taxonomic analysis of asteroids. The motivation of this type of research is to explain the initial composition of asteroids throughout the Solar System \citep{De15}. The taxonomic classification is a proxy to the mineralogy of the object and, consequently, to its spectrum. Over the years, as each survey provided information at different wavelengths, several taxonomic classifications have been developed, from \cite{Th84} to \cite{D09}, and spanning different taxonomical classes. However, we can determine that the essential taxonomical types are: the C-complex (spectrum similar to carbonaceous chondrites, albedos less than 0.1, very dark objects), the S-complex (albedo of typically 0.2, characterized by spectra with moderate silicate absorption features at 1 and 2 $\mu$m), and the X-complex (characterized by moderately sloped and subtly featured or featureless spectra, \citealt{De15}). In addition, there are spectral types that do not fit into these three main complexes. Among them are the types A, D, V, K, P, Q, L. We note that in this work we use the terms "complex" or "taxonomies" to refer to a given taxonomic type and "complexes" or "taxa" for the plural. 

Previous efforts in this regard using Sloan data and novel methodologies include \cite{C10} and \cite{S21}, who used probabilistic methods to assign taxa, while \cite{H16} and \cite{R22} used machine learning techniques. They all have their criteria to handle objects with multiply observations and quality cuts. In the case of objects with multiple observations, the nominal values are usually an average over different measurements or, at worst, a single epoch measurement; both cases usually neglect phase angle effects. In \citet[AC22]{Al21}, we computed absolute magnitudes in all filters for objects in the MOC (for details, see AC22). In short, we used the HG$^*_{12}$ model to obtain $H$ and $G^*_{12}$ for almost 14\,000 objects, about 85 \% in all five filters. Moreover, Fig. 16 of AC22 shows differences when using a color that has not been corrected by phase effects and that it may impact the taxonomical classification of the objects.

Therefore, this work combines the absolute magnitudes obtained in different filters and taxonomic classifications. This idea stems from the fact that the taxon of an asteroid can be mistaken if the applied colors are affected by phase-angle effects. We consider that a classification based on absolute magnitudes represents an improvement in taxonomic classifications. Furthermore, we used machine learning (ML) tools for the analysis. It is well known that this type of technique has proven to be very efficient in recent years, and it is and will surely be one of the most frequently applied methods in the era of large observational surveys. 

In Section \ref{2}, we present the databases used for the present work. In Section \ref{3}, we introduce the basic concepts of ML and the tools used for our analysis. Finally, in Section \ref{4}, we present our conclusions.

\section{Dataset}\label{2}

We used the absolutes magnitudes computed by AC22. They worked with the fourth release of the Sloan Moving Objects catalog (hereafter just MOC, \citealp{I01, J02}) and increased this database with the updated SVOMOC version \citep{C16}. The authors estimated the absolute magnitudes of the objects in the five Sloan filters (u', g', r', i' and, z') using the HG$^*_{12}$ photometric model, applying a method that combines Monte Carlo simulations and Bayesian inference to include the effect of rotational variations. The final catalog contains absolute magnitudes for almost 15\,000 objects, with about 12\,000 of them including all five filters. The authors noted that several asteroids with multiple observations led to ambiguous classification. We can address this issue using absolute magnitudes because phase-angle color variations do not affect them. On the other hand, we used the \citet[C10]{C10} SDSS-based taxonomic classification to compare our results. 

We used a first data set of 5168 asteroids for a preliminary analysis and phase space exploration. We selected these objects based on the following: 1) we kept only asteroids with $H$ in all five Sloan filters; 2) we removed those objects with ambiguous classification according to C10; and 3) we kept objects with uncertainties < 0.4 mags, since in preliminary calculations, we noted that those asteroids whose absolute magnitude uncertainties exceed 0.4 mags may result in unreliable taxonomic classifications. We note that, as explained in Section \ref{phasespace}, to run the clustering algorithm, we use a larger sample of 9481 asteroids that have observations in magnitudes $H_g$, $H_i$, and $H_z$, and no ambiguous classification according to C10. Throughout this work, we use $u$, $g$, $r$, $i$, and $z$ to refer to the apparent Sloan magnitudes and $H_u$, $H_g$, $H_r$, $H_i$, and $H_z$ to refer to the absolute magnitudes calculated by AC22.

\section{Unsupervised machine learning}\label{3}

Machine learning is a branch of computer science that attempts to create algorithms capable of learning about a specific problem from a set of data without specific programming. Different studies of asteroids use Machine Learning tools; for instance, \cite{Ca19}, \cite{Cam19}, \cite{Pe21}, \cite{Ca21}, \cite{Dela21}. \cite{Ca21} all present an excellent review of the topic.

Machine learning algorithms are classified according to multiple criteria. The most common one that divides them into supervised and unsupervised algorithms. For the present work, we concentrate on the second group. Unsupervised algorithms {start} from an unlabeled data set. These algorithms aim to analyze the training data set and extract relevant information. The method we apply for our analysis is clustering. 

\subsection{Phase space exploration}\label{phasespace}
\begin{figure}
    \centering
    \includegraphics[width=\columnwidth]{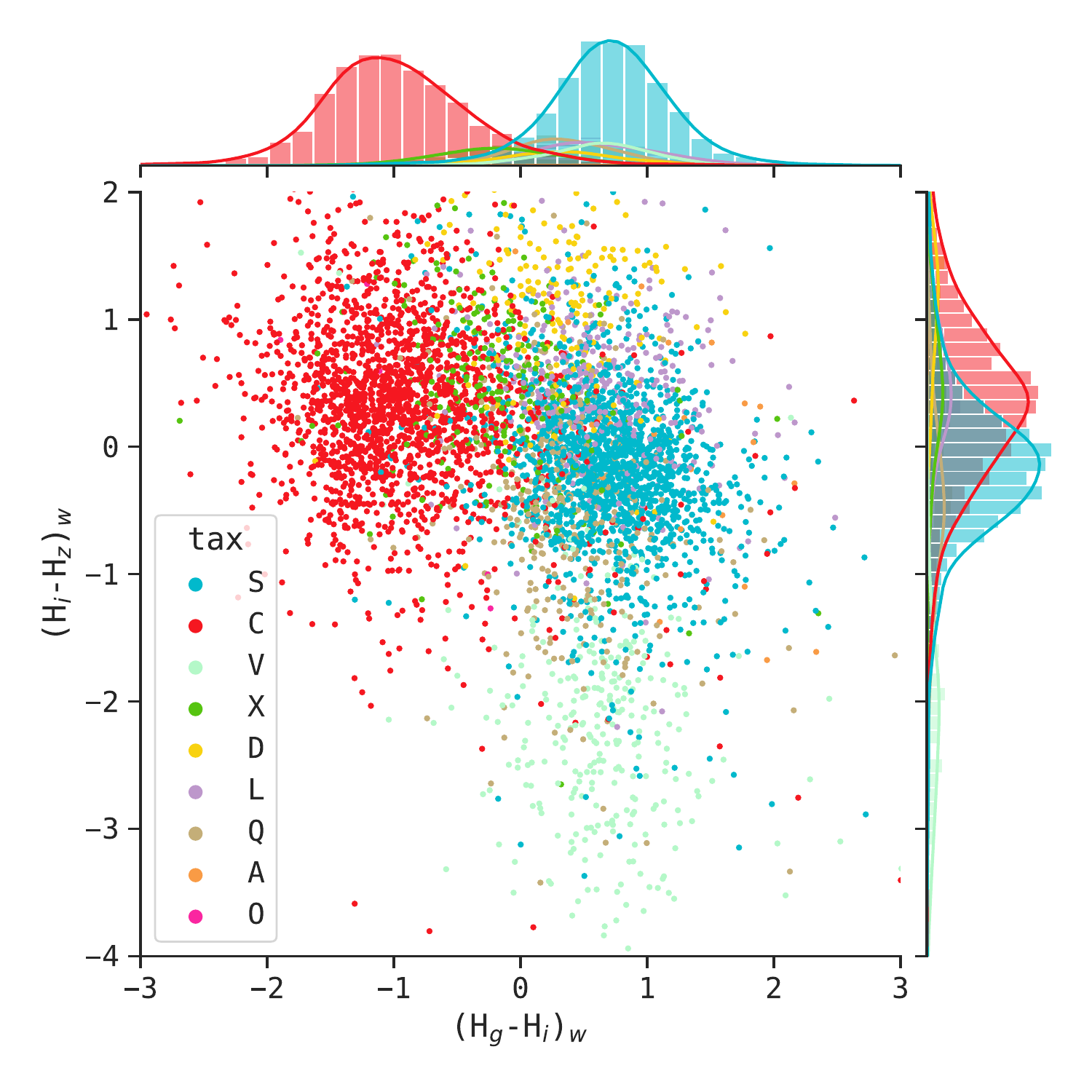}
    \caption{Color-color diagram in absolute magnitudes for 5168 asteroids of our sample. Color scheme corresponds to C10's Taxonomy Classification.}
     \label{color-color}
\end{figure}

Our first step is to determine the parameters that best describe the set of observations. The objective is to find two variables that provide a good division of the sample into subgroups, to identify the different taxa. We calculate the colors ($H_i - H_j$) using the five Sloan filters. Then, we apply whitening to analyze the relationships between them. A  ''white'' variable $X_w$ is defined as $$X_w=\frac{X-\overline{X}}{\sigma_X},$$ where $\overline{X}$ is the average of the original variable and ${\sigma_X}$ is its original standard deviation. By construction, a white variable is dimensionless with a mean value of zero and a standard deviation equal to one. In Fig. \ref{parameters} we visualize the correlations obtained, including the five whitened colors. The figure shows that the parameters that best separate the sample are the colors $(H_g-H_i)_w$ versus $(H_i-H_z)_w$. Figure \ref{color-color} shows the resulting distribution of points. On the x axis, there are well differentiated clusters corresponding to complexes C and S. On the y axis, the division of the V-complex asteroids is distinguished. Originally, we also included the whitened albedo $p_w$, but it did not provide any additional information; therefore, we dropped it from the analysis and  from Fig. \ref{parameters}.

\begin{figure}
    \centering
    \includegraphics[width=\columnwidth]{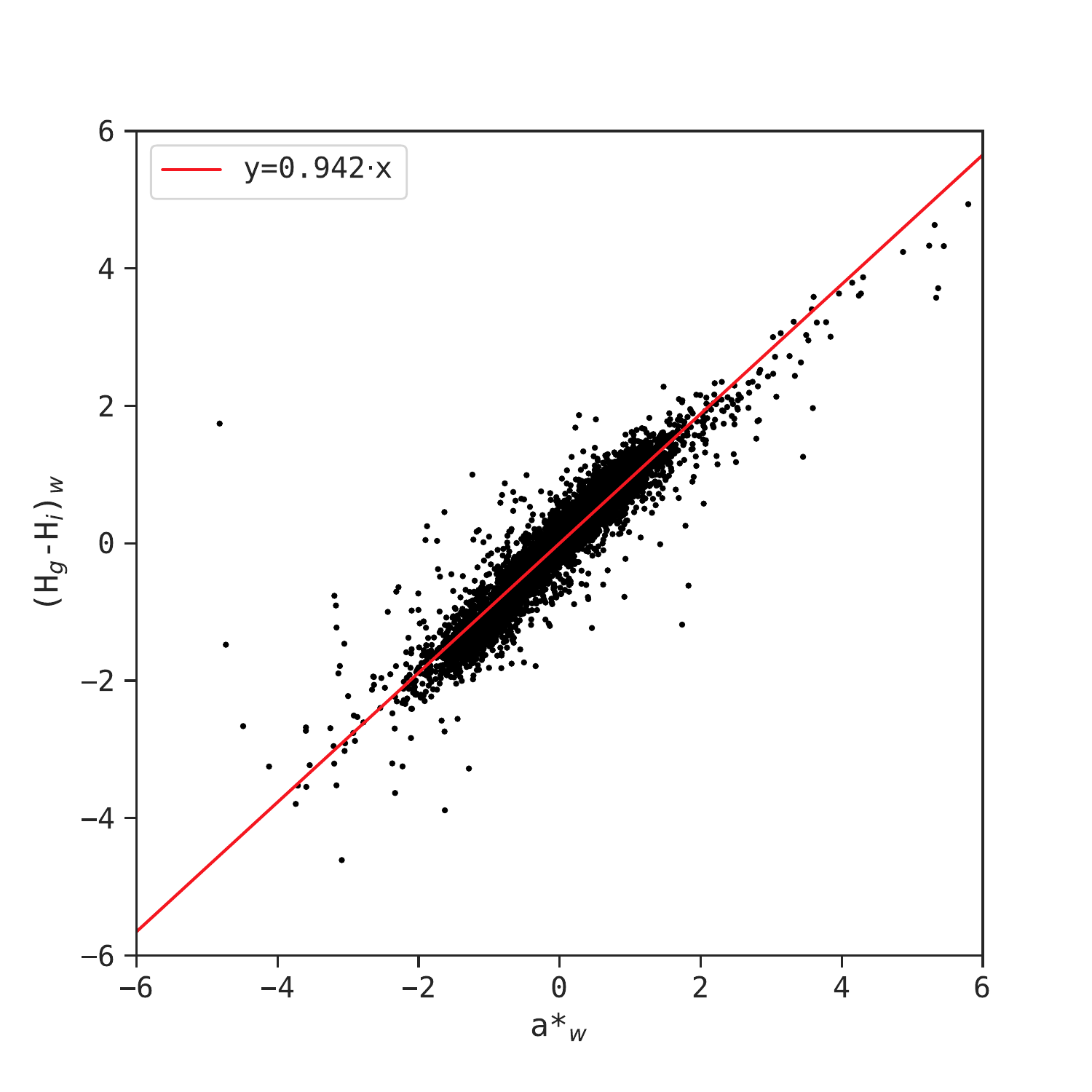}
    \caption{a*$_w$ vs $(H_g-H_i)_w$ linear correlation. }
     \label{linear}
\end{figure}

Many authors (\citealp{I01, B02, R06, D09}) implemented PCA (Principal Component Analysis) to obtain a different color known as a*. To obtain it, one applies PCA to the space $g-r$ versus $r-i$, where a* is the first component. We calculate a* and visualize the a*$_w$ versus $(H_i-H_z)_w$ space to compare the obtained distributions. We note that the obtained distribution is very similar to that of Fig. \ref{color-color}. A more direct way to observe this correspondence between graphs is to look at the relationship between the parameters a*$_w$ and $(H_g-H_i)_w$  in Fig. \ref{linear}. The correlation coefficient between both parameters is 0.942.

To calculate a*$_w$ and plot it against $(H_g-H_i)_w$, we need to know four magnitudes of the object ($H_g$, $H_r$, $H_i$, and $H_z$) and apply an unsupervised machine learning (PCA) method. However, if we use the $(H_g-H_i)_w$ versus $(H_i-H_z)_w$ space, we only need three magnitudes ($H_g$, $H_i$, and $H_z$) and no extra computation. Therefore, we decided to apply clustering to the $(H_g-H_i)_w$ versus $(H_i-H_z)_w$ space. Using as few magnitudes as possible allows us to work with larger samples because, in the AC22 catalog, there are objects with missing $H$ in some filters. 

After this preliminary exploration of the phase space, we decided to no longer use the sample of 5168 asteroids. In Section \ref{taxclass}, we use a sample of 9481 objects consisting of 1) asteroids that have measured $H_g$, $H_i$, and $H_z$; and 2) have no ambiguous classification according to C10. 

\subsection{Taxonomy classification}\label{taxclass}

As mentioned before, clustering is an unsupervised machine learning algorithm. It divides the given data into different clusters based on their distances. We identify two different methods of clustering: Hard clustering, each datum belongs to a cluster completely or not (e.g., the K-means algorithm), and soft clustering, a datum may belong to more than one cluster with some probability or likelihood value (e.g., fuzzy C-means).

Since we have overlapped data, it is more convenient to apply fuzzy C-means. We present results obtained using {\tt skfuzzy.cmeans} from Scikit-Fuzzy (\url{10.5281/zenodo.3541386}), a python implementation. For a detailed description of the algorithm, see \cite{R10}. Nevertheless, we present a very brief summary here. Supposing we have a set of $n$ data samples (in this case $n = 9481$), then:
\begin{equation}
    X = \left\lbrace \vec{x}_1,\vec{x}_2,...,\vec{x}_n \right\rbrace.
\end{equation}
Each data sample, ${x}_j$, is defined by $l$ features, (in this case $l = 2$, corresponding to $(H_g-H_i)_w$ and $(H_i-H_z)_w$):
\begin{equation}
    \vec{x}_j = \left\lbrace {x}_{j1},{x}_{j2},...,{x}_{jl} \right\rbrace.
\end{equation}
We define a family of sets $\left\lbrace A_i, i=1,...,c \right\rbrace$ where $c$ is the number of clusters.
The objective function of this optimization problem is:
\begin{equation}
    J_{l}(\textbf{U}, \textbf{v}) = \sum_{k=1}^{n}\sum_{i=1}^{c} (\mu_{ik})^m (d_{ik})^2,
\end{equation}
where:
\begin{equation}
    d_{ik} = d(\vec{x}_{k} - \vec{v}_{i}) = \sqrt{\sum_{j=1}^{l} (x_{kj}- v_{ij})^2},
\end{equation}
and \textbf{U} is the fuzzy partition matrix, which contains the probability of each point $\vec{x}_n$ to belong to each of the $c$ clusters; $\mu_{ik}$ is the membership of the $k^{\rm th}$ data point in the $i^{\rm th}$ cluster; the parameter $m$ is the fuzziness parameter (generally taken as 2, \cite{PB95}). This value has a range $m \in [1,\infty)$ and it controls the amount of fuzziness in the classification process.

Instead of minimizing the objective function $J_l$, the algorithm applies an iterative optimization (Bezdek, 1981) according to the following steps: 1) Select the $c$ number of clusters and the value of parameter $m$. Initialize the partition matrix $\textbf{U}^{(0)}$; 2) Compute the centers $\vec{v}$ of each cluster; 3) Update the partition matrix, $\textbf{U}^{(1)}$; 4) Calculate the norm of the difference between the two matrices $\textbf{U}^{(0)}$, $\textbf{U}^{(1)}$. Verify if it is less than the prescribed level of accuracy, $\epsilon$, to determine whether the solution is good enough, that is: $\parallel \textbf{U}^{(1)} -  \textbf{U}^{(0)} \parallel \leq \epsilon$. If this condition is satisfied, stop. Otherwise, return to step 2.

\begin{figure}
    \centering
    \includegraphics[width=\columnwidth]{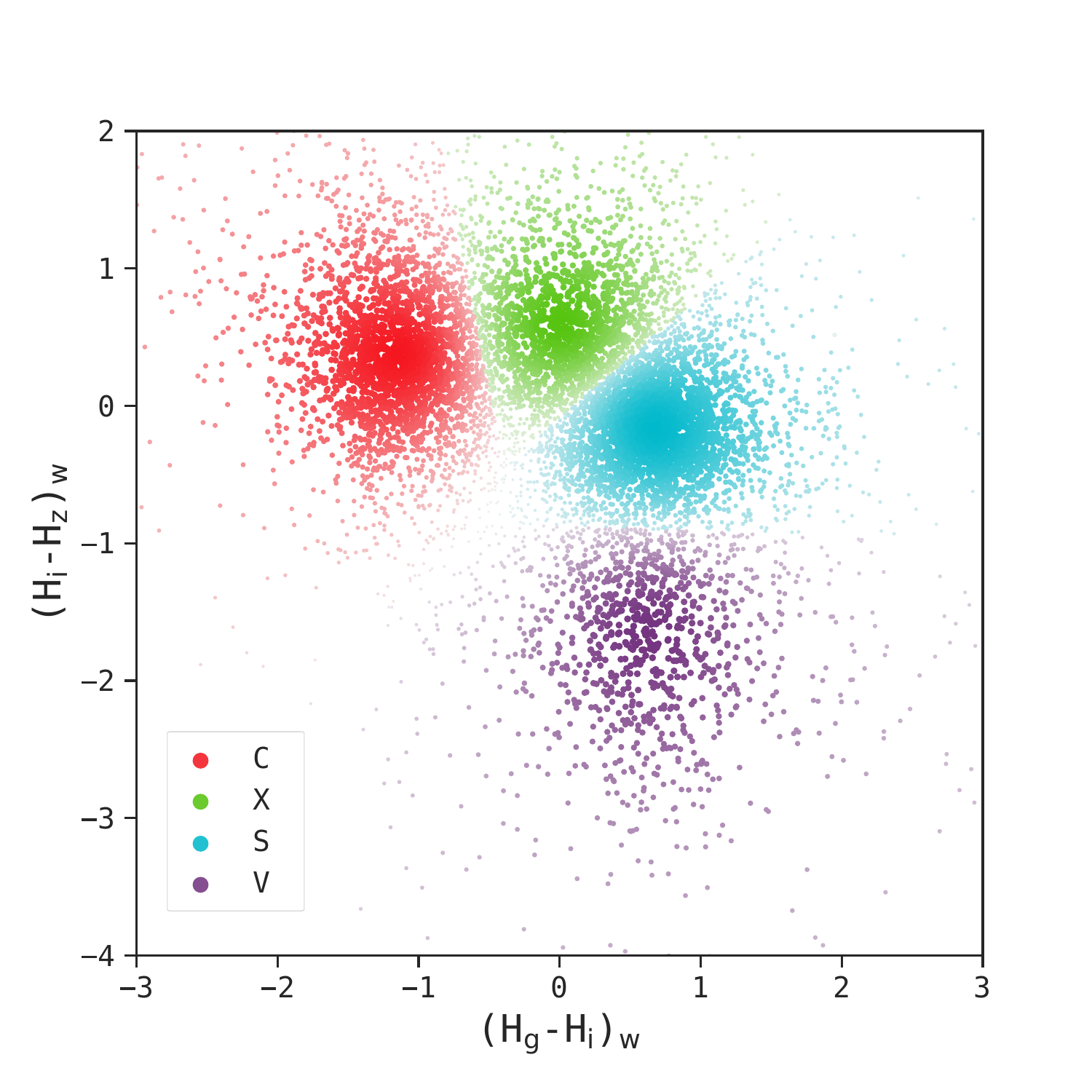}
    \caption{Identification of four taxonomic groups in our sample of 9481 asteroids using the fuzzy C-means algorithm. The red cluster corresponds to C-complex, the green cluster corresponds to X-complex, the blue cluster corresponds to S-complex, and the violet cluster corresponds to V-complex. The probability of each point belonging to a given cluster is proportional to the intensity of the color.}
     \label{prob}
\end{figure}

This work aims to distinguish the four most relevant taxonomic complexes: C, S, X, and V. For this reason, we built a four-cluster model for prediction, generating new uniform data, and predicting belongings. In Fig. \ref{prob} we can see the four clusters identified with different colors. In this graph, the probability is proportional to the intensity of the color. 

A quick comparison between Figs. \ref{color-color} and \ref{prob} allows us to distinguish the correspondences of each cluster: the red cluster corresponds to C-complex, the green cluster corresponds to X-complex, the blue cluster corresponds to S-complex, and the violet cluster corresponds to V-complex. We note a good separation between the C, S, and V complexes. From the total of 9481 asteroids in the sample, we selected as good classifications those with a probability greater than 60$\%$. We find asteroids with probabilities lower than 60$\%$ at the edges of the clusters, so their final classification is ambiguous. We classified 6181 asteroids following our criteria. A comparison with C10 finds 4074 objects with the same taxon, while 1016 asteroids present new taxa. For 1091 objects, the taxa determined in C10 (who separated into nine classes) is not included in our four complexes.

\begin{figure}
    \centering
    \includegraphics[width=\columnwidth]{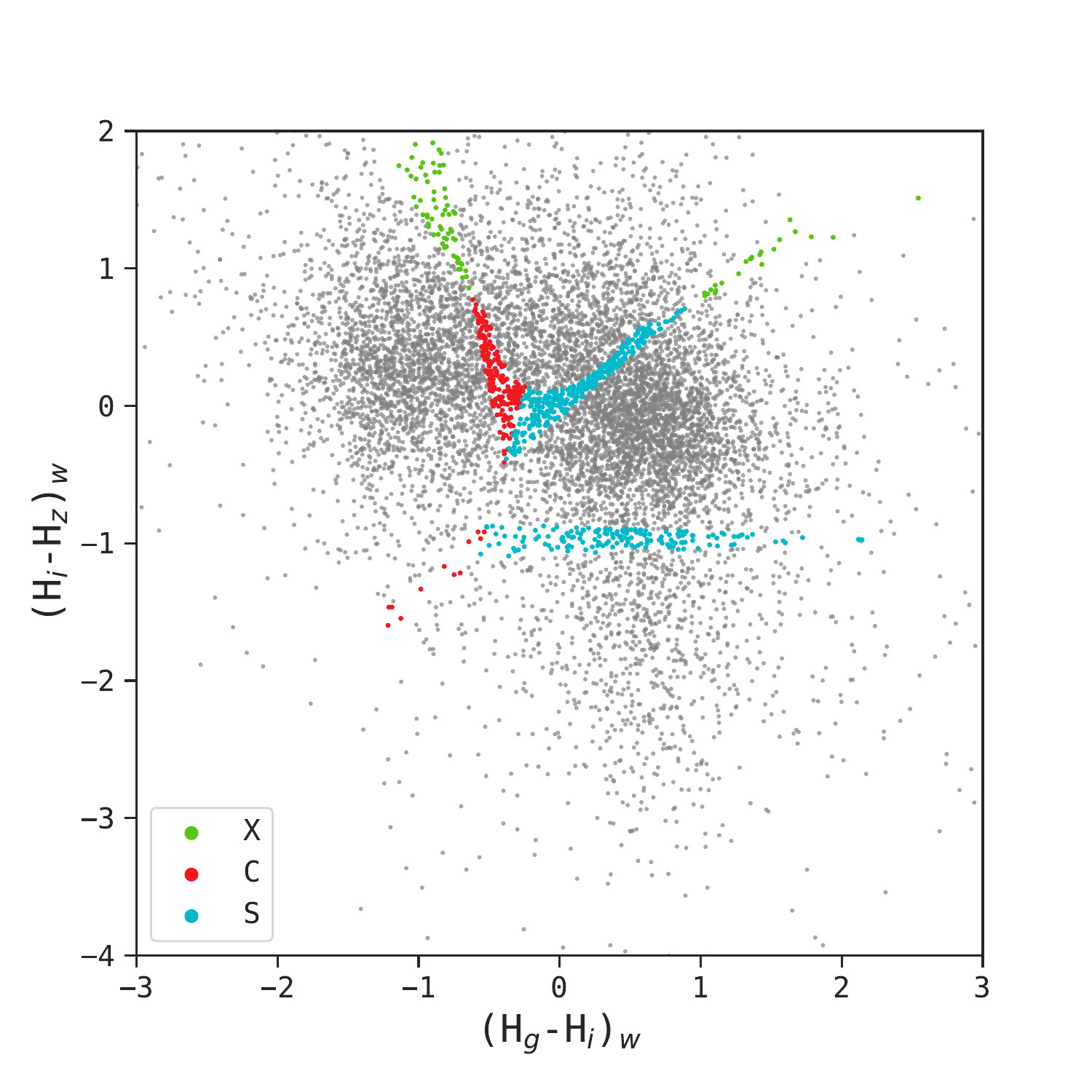}
    \caption{Identification of cluster membership by K-means for asteroids with membership probabilities lower than 60$\%$.}
     \label{k-c}
\end{figure}

Since with K-means is the more common method, we compare the outcome of the two. To do so, we analyzed the sample of 3301 asteroids with classification probability less than 60$\%$ according to the C-means algorithm. We found that 2494 asteroids were assigned to the same taxonomic type by applying the K-means algorithm, while 807 were grouped into a different cluster. As expected, classification differences occur in the overlapping regions of the clusters, especially at the edges of taxonomic group X (Fig. \ref{k-c}).

\subsection{Reclassification of ambiguous targets}

\begin{figure*}
    \centering
    \includegraphics[width=\textwidth]{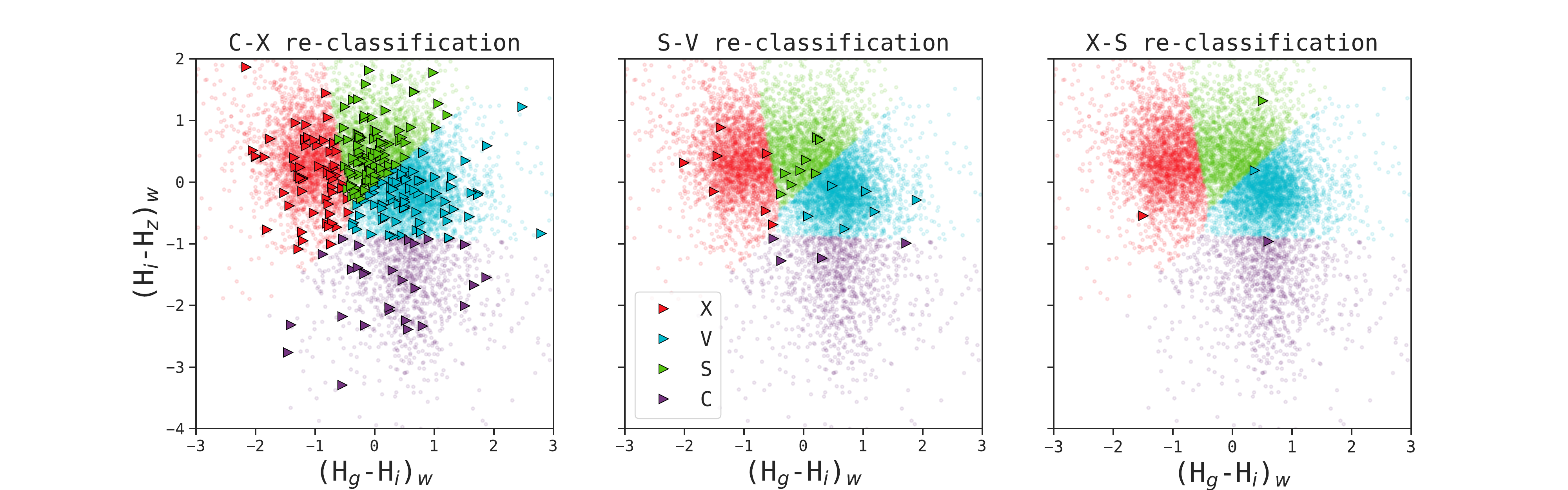}
    \caption{Reclassification of objects with ambiguous taxonomy C-X, S-V, and X-S according to \cite{C10}. The triangles represent the objects with ambiguous classification in C10 and where they are in our scheme. The color scheme is the same as in Figure \ref{prob}}
     \label{reclass}
\end{figure*}

\cite{C10} followed a scheme that classifies some asteroids with two different taxa. One problem that the authors exposed about the SDSS MOC4 is that it includes several asteroids with multiple observations, which increases the possibility of having multiple classifications. We resolved this fact using absolute magnitudes. 
We present the reclassification of those asteroids that present ambiguities between the C, S, X, and V complexes. For this purpose, we used the model constructed above. In Fig. \ref{reclass}, we show the reclassification of ambiguous C-X (left panel), S-V (center panel), and X-S (right panel) asteroids. 

In the case of C-X asteroids, we reclassified 134 objects (out of 270) with a probability higher than 60$\%$ in our scheme. Although C10 classifies them as type C or X, these asteroids are widely scattered in our parameter space, showing that they belong to different taxa. As seen in the left panel of Fig. \ref{reclass}, we classified as V-complex (in violet) several asteroids that were ambiguous C-X. We checked case by case if these objects present the decrease in flux at the z filter, and we present the results in Sect. 3.5. We did the same with the ambiguous S-V in C10, and we reclassified 12 out of 26 objects, mainly in our S and X space. We show the results in the center panel of Fig. \ref{reclass}.
The right panel of the figure shows the reclassification of ambiguous X-S asteroids. We reclassified two out of four with a probability higher than 60$\%$.

\subsection{Orbital distribution}
\begin{figure}
    \centering
    \includegraphics[width=\columnwidth]{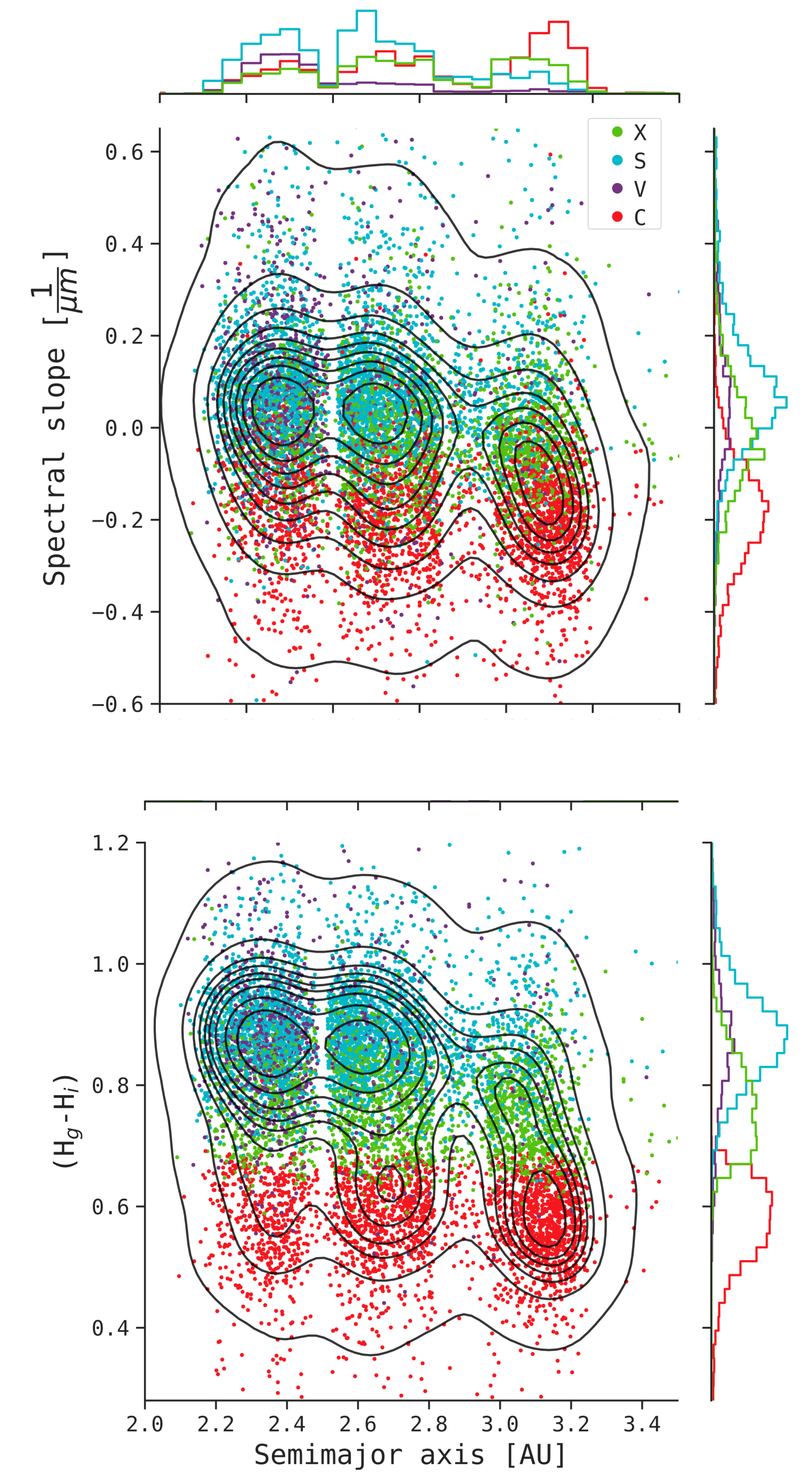}
    \caption{Semimajor axis distribution of asteroids with more than 60$\%$ probability of belonging to the assigned taxonomic class). In panel a), we show semimajor axis vs Spectral slope. In panel b), we present semimajor axis vs $H_g-H_i$. For both, on marginal axes we show histograms for each spectral class. We also plot the contours of the 2D density. Color scheme for spectral type are the same as before.}
     \label{distribution}
\end{figure}

We analyzed the distribution of taxonomic complexes in the semimajor axis, $a$ versus spectral slope (Fig \ref{distribution}a) and semimajor axis, $a$ vs $H_g-H_i$ planes (Fig \ref{distribution}b). We extracted the semimajor axis from the Astorb database\footnote{ \url{https://asteroid.lowell.edu/main/astorb/}}. Our sample includes the 6181 asteroids selected from the first clustering process plus the 148 reclassified asteroids, all with classification probability greater than 60$\%$. To obtain the spectral slope, we calculate the relative reflectances as in \cite{Al13}:
\begin{equation}
    F(\lambda)=\{F_{j}\}=\{10^{-0.4[(H_j-H_g)-(j-g)_\odot]}\},
    \label{fluxeq}
\end{equation}
where $j$ represents the $u$, $r$, and $i$ magnitudes, and $\odot$ represents the Solar colors\footnote{\url{https://www.sdss.org/dr12/algorithms/ugrizvegasun/}}. We use the g filter as a reference for consistency with C10's work. We smooth the obtained spectra with a linear spline and fit it with the least squares function $F(\lambda)=slope\times\lambda+b$. 

The marginal chart on the top of Fig \ref{distribution} shows the relative frequency of objects per taxa over the range of semimajor axes in the Main Belt. We obtained the expected distribution of taxa according to current constraints about the Solar System. The S-complex predominate in the inner Main Belt, and the C-complex asteroids dominate the outer region, as was found in \citep{De14}. In addition, it is possible to note the peak of the V-complex asteroid in the Vesta family region. The X-complex asteroid peak at 3 AU, was also previously observed by other authors \citep{Mo03,Fo11}. 

On the other hand, the distribution of spectral slopes shows a decrease in the slope going toward the outer Main Belt. This is because redder sloped objects (S or V complex) dominate the inner region, while C-complex, which tend to be neutral or slightly bluish, dominate the outer region. This may seem at odds with the classical view that redder spectra appear in the outer belt (D or P taxa), but, first, we do not discriminate these two, which are probably assimilated into the C or X complexes, and there is a clear decrease in the number density of the objects in the outer regions, which explains our results.
As we can see in the marginal histograms on the left of Fig. \ref{distribution}a the mean spectral slope of the four complexes are (C,S,X,V) $= (-0.185, 0.075, 0.049, -0.015)$ $\mu$m$^{-1}$, while the width of the histogram, represented by its standard deviation, is (C,S,X,V) $= (0.11, 0.12, 0.16, 0.12)$ $\mu$m$^{-1}$. The only wider histogram is that of the V-type asteroids, and this may be due to an influence of the absorption band of pyroxene at 0.9 $\mu$m on the i filter magnitude. Depending on the surface composition of the asteroid, the 0.9 $\mu$m absorption band may be deeper or may have its center moved towards 0.93 $\mu$m \citep{Duffard2006}. Thus, the composition of the pyroxenes present in the V-type asteroids would influence the spectral slope distribution of the V-complex asteroids, while in the other complexes, C, X, and S, the standard deviation is similar between them, around 0.12 $\mu$m$^{-1}$. We also note that the histogram corresponding to the C complex contains many negative values, which may be related to B-type asteroids in this sample, that are not differentiated from the C-complex because we only took four spectral classes. B-type spectra are featureless with a negative slope, meaning bluer spectra. The right panel of Fig \ref{distribution}b offers clear evidence of how well the color $H_g-H_i$ separates the C and S taxa.

\subsection{V-complex asteroids}
\begin{figure}
    \centering
    \includegraphics[width=\columnwidth]{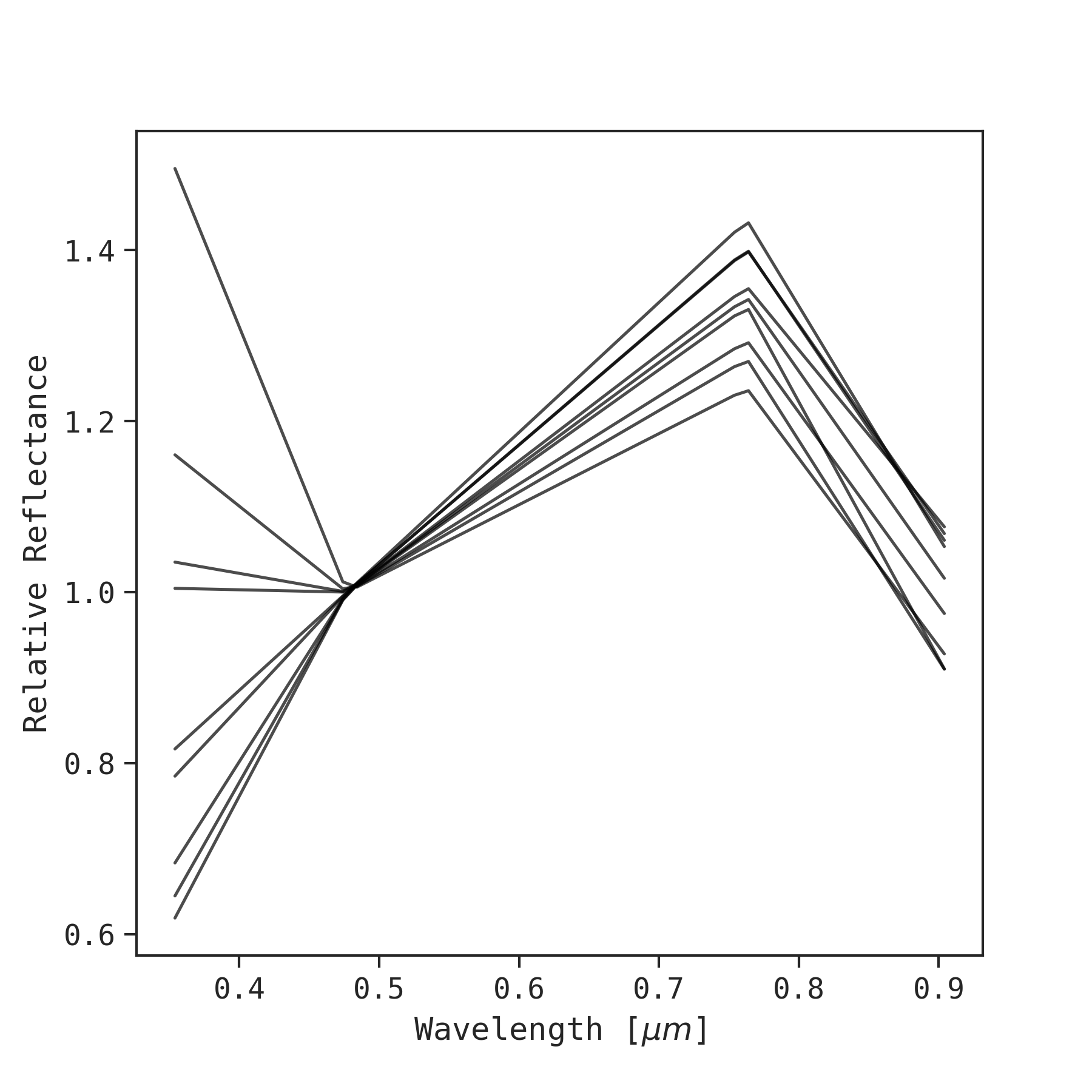}
    \caption{Spectrum of V asteroids outside the Vesta family.}
     \label{spectrum}
\end{figure}

Another interesting feature seen in Fig. \ref{distribution} is the presence of V-complex asteroids outside the Vesta family region. We studied some of these cases in particular. We isolated those objects whose semimajor axis is greater than 2.8 AU \citep{duffard2009}, which is the inner limit of the outer Main Belt, and with a classification probability greater than 80$\%$. We calculate the reflectances in each filter according to Eq. \ref{fluxeq}. In Fig. \ref{spectrum} we show the obtained photospectra to confirm the V-complex spectrum of each candidate. It is important to note that they all have the characteristic peak at $0.75$ $\mu$m and an absorption band near $0.95$ $\mu$m. \cite{R06} used the third release of the MOC to find five possible V-type asteroids in the outer belt: (7472) Kumakiri, (10537) 1991 RY16, (44496) 1998 XM5, (55613) 2002 TY49, and (105041) 2000 KO41. \cite{duffard2009}  spectroscopically confirmed two of them: (10537) 1991 RY16 and (7472) Kumakiri. Our list of V-complex candidates in the outer belt is: (100105) 1993 FK35, (86162) 1999 RR205, (97502) 2000 CL93, (105353) 2000 QN105, (39069) 2000 VM10, (135160) 2001 QL241, (111422) 2001 XM196, (160213) 2002 CS252, (338278) 2002 TQ327, (143194) 2002 XT84, (197503) 2004 BN97, (171258) 2005 QW84, (261001) 2005 SN95, (242740) 2005 UP476, and (224611) 2005 YV63.

{On the other hand, \cite{H14} identified seven new V-type asteroid candidates in the outer main belt: (11465) 1981 EP30, (55270) 1998 QA73, (91159) 1999 XG203, (34698)  2001 OD22, (92182) 2001 PR10, (208324) 2001 RT147, and (177904) 2005 SV5, using MOC gri slope and $z' - i'$ colors to classify an asteroid depending its position on that
plane. Also,} \cite{licandro2017} identified new basaltic V-type asteroids using near infrared colors observed by the VHS-VISTA survey and compiled them into the MOVIS-C catalog. Even though they used a different catalog, the method is similar to ours: they filtered the candidates with specific criteria on the infrared colors. More recently, \cite{miglio2021} presented new spectral observations of 23 putative V-type asteroids selected according to color surveys in the visible from the MOC and near infrared from the MOVIS catalog. None of the objects we present are reported in Licandro's nor Migliorini's respective works.

Identifying basaltic material is important because some V-complex asteroids discovered far from the Vesta family support the hypothesis of multiple progenitors of basaltic asteroids in the early Solar System. In the main belt, Eunomia has been suggested as a possible parental body \citep[and references therein]{Ca07, Ca14}. Furthermore, for the outer main belt, \cite{H14} proposed Eos and Koronis as parental bodies. We want to stress that V-complex asteroids are relatively easy to identify, both spectroscopically and photometrically, due to their unique spectral behavior. The advantage of the photometric method is the ability to reach a fainter population in less time than strictly using  spectra.

\subsection{Dynamical families}
\begin{figure}
    \centering
    \includegraphics[width=\columnwidth]{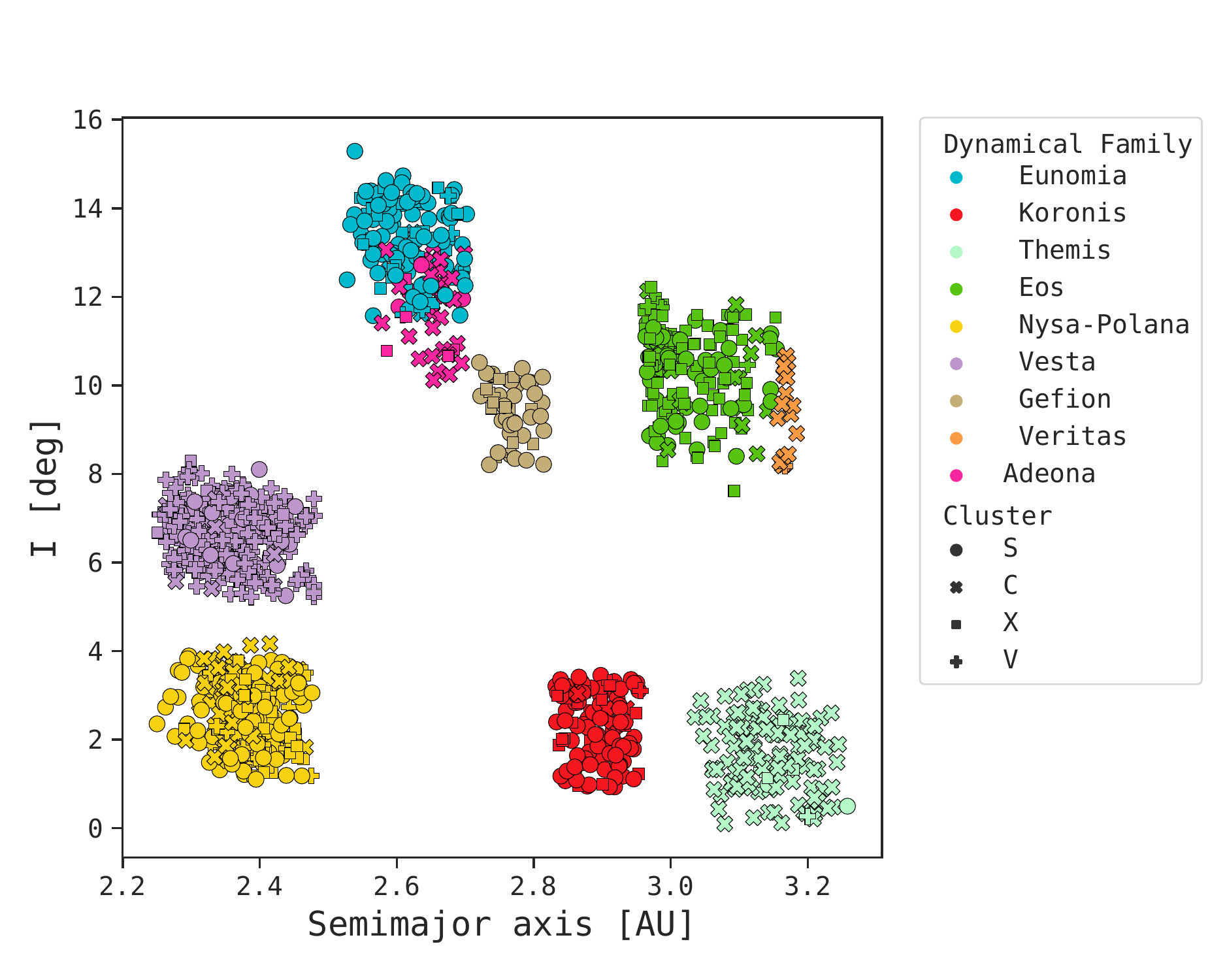}
    \caption{Identification of dynamical families in our sample. Different colors indicate different families. We also distinguish the spectral types with markers. }
     \label{fam}
\end{figure}

We were able to identify 1253 asteroids belonging to dynamical families according to \cite{Ne10}. For this task, we matched our sample with Nesvorny's catalog \citep{Ne20}. In Fig. \ref{fam} we show our results. We chose to restrict our analysis to the larger families (Vesta and Nysa in 2.4 AU, Eunomia and Adeona in 2.6 AU, Gefion in 2.8 AU, Koronis in 2.9 AU, Eos in 3.0 AU, and finally Themis, and Veritas in 3.2 AU). In Fig. \ref{fam}, we show the family members and the respective taxa. Several works have been devoted to investigating the composition of asteroid families using photometric catalogs \citep{licandro2017,wong2017,morate2018}. 
We can see a mixture of X and S complexes in the Koronis, Eos, and Eunomia families. Then, in the outer part of the belt, all the families belong to the C complex (Themis and Veritas) and Adeona in 2.6 AU. The region of the Nysa complex is a mixture of the C and S taxa, showing the heterogeneity that reigns there: the actual Nysa family composed of S-complex and the primitive Polana and Eulalia families \citep{walsh2013,deleon2016,pinilla2016}.

We know that asteroids belonging to the same family should share the same taxon because they all come from the same parental body. We have reproduced this behavior with our results except for two families: Eos and Koronis, because of the difficulty in unambiguously discerning the X and S complexes. We attempted tests using albedo values measured and published by AKARI to address this problem. However, we do not find any separation of the groups using the albedo. Moreover, the sample is significantly reduced due to the few albedo measurements currently available. Looking at a graph that is analogous to Fig. \ref{color-color}, we notice that the families do not have a particular distribution in this parameter space. Such exceptions have already been discussed in the literature before, for instance, in \cite{Ca13} where asteroid families are identified in domains of proper elements, albedos, and SDSS colors, and Eos and Koronis were found to be not as homogeneous as other families obtained with this approach. According to the authors, this may be caused by the possible differentiated, or partially differentiated, nature of the parent body of these families.

The final results are shown in Table \ref{table:1} below. The uncertainties of colors H$_g$-H$_i$ and  H$_i$-H$_z$ were calculated with the classical error propagation method. The error of the slope is provided by the linear adjustment implementation used. In some cases, it was not possible to calculate the slope since there was no information on the 4 values of Flux. In addition, not all asteroids were assigned to a family. In both cases, we indicate missing values with '-'. 
\begin{table*}
\caption{Sample of the results}
\label{table:1}
\centering
\begin{tabular}{c c c c c c c c c c}
\hline\hline
ID   & H$_g$-H$_i$ & $\sigma_{H_g-H_i}$ & H$_i$-H$_z$ & $\sigma_{H_i-H_z}$ & Spectral slope ($\frac{1}{\mu m}$) & $\sigma_{Slope}$ ($\frac{1}{\mu m}$) & Tax & Prob & Fam \\
\hline
1220T-2 & 0.586052 & 0.369842 & -0.034118 & 0.363174 & -0.214953 & 0.00485631 & C & 0.814686 & Themis \\
1977EW7 & 0.893813 & 0.425204 & -0.020652 & 0.397426 & -0.0648163 & 0.0423916 & S & 0.926829 & Eunomia \\
1978VN4 & 0.874879 & 0.490940 & 0.011894 & 0.601412 & - & - & S & 0.711725 & Massalia \\
1013T-2 & 0.799885 & 0.412935 & 0.039389 & 0.375412 & 0.375412 & 0.00485375 & X & 0.837800 & - \\
2001SV291 & 0.591414 & 0.329793 & -0.083079 & 0.297460 & -0.205003 & 0.00100782 & C & 0.406850 & Adeona \\
\hline
\end{tabular}
\tablefoot{The first column indicates the object's ID, the second the value of H$_g$-H$_i$, the third shows the uncertainty of H$_g$-H$_i$, the fourth the value of H$_i$-H$_z$ and the fifth its uncertainty. Then, the sixth column presents the value of spectral slope, and the seventh its uncertainty. The last three columns give the taxonomic classification obtained in this work, the probability of classification, and the family to which the object belongs. The complete table can be obtained upon request. It contains 9781 asteroids, since it does not have the classification probability threshold over 60\%. The complete catalog is available at the CDS (https://cdsarc.u-strasbg.fr/) or upon request.}
\end{table*}

\section{Conclusions}\label{4}
We determined the taxonomic classification of 6329 objects using absolute colors computed by AC22 and ML methods, which are particularly fuzzy C-means. Using absolute colors frees us from possible ambiguities that naturally arise when using databases with multi epoch observations.

Instead of using the (many) different taxonomies used in the literature, we chose to restrict to four main complexes: C, S, X, and V, because they encompass most of the known spectral behavior of asteroids. Within each complex, it is possible to parametrize the behavior using simple quantities such as the slope of the photo-spectra and the absolute color $(H_i-H_z)$.

The distributions of spectral slopes within each complex show a broad mix of objects and no clear pattern. These distributions suggest that not necessarily redder objects appear at higher semimajor axes. On the other hand, the distribution of complexes along the main belt of asteroids does not uncover any new feature. The S-complex dominates the inner and middle belt, and the C-complex dominates the outer belt. The V-complex appears concentrated in the inner belt, coincident with the location of the Vesta family, but, interestingly, there are many candidates in the middle and outer belt. These objects deserve further spectroscopy observations for confirmation.

Spectroscopic surveys of asteroid families feature large samples of spectra for each one, but using absolute photometric techniques, we can reach even more objects not affected by phase coloring. Furthermore, to unveil an as-yet-undiscovered faint population of objects, we can prepare our software skills for the next photometric surveys, such as Gaia DR3, JPLUS, LSST, DES, and more. 

By applying the identification of asteroid families to the MOC, we can gain knowledge of the completeness of our spectroscopic survey of each family. In addition, we can test crossmatching techniques using different databases. One of the main results of this paper is the development of a technique to obtain absolute photometric spectra of asteroids by using photometry in several filters. We need to highlight the importance of crossmatching databases (family clustering identification, orbital data, JPL ephemeris, albedo from AKARI \citealp{AKARI}, WISE \citealp{WISE}, NEOWISE \citealp{NEOWISE}, or IRAS \citealp{IRAS}, as well as those observed by the telescopes).

\begin{acknowledgements}
      M. C is a doctoral fellow of CONICET (Argentina).
      
      Funding from Spanish projects PID2020-112789GB-I00 from AEI and Proyecto de Excelencia de la Junta de Andaluc\'ia PY20-01309 is acknowledged.
      
      R.D and A.A.C acknowledges financial support from the State Agency for Research of the Spanish MCIU through the "Center of Excellence Severo Ochoa" award to the Instituto de Astrofísica de Andalucía (SEV-2017-0709).
      
      R.D acknowledges funding from National Spanish project PID2020-112789GB-I00 and the regional Junta de Andalucía PY20-01309. 
      
      This work used https://www.python.org/, https://www.scipy.org/, Matplotlib \citep{h07} and Scikit-learn \citep{P11}.
\end{acknowledgements}

%
%

\begin{appendix} 
\section{Correlation between absolute colors}
    \begin{figure}[h!]
    \onecolumn\includegraphics[width=\textwidth]{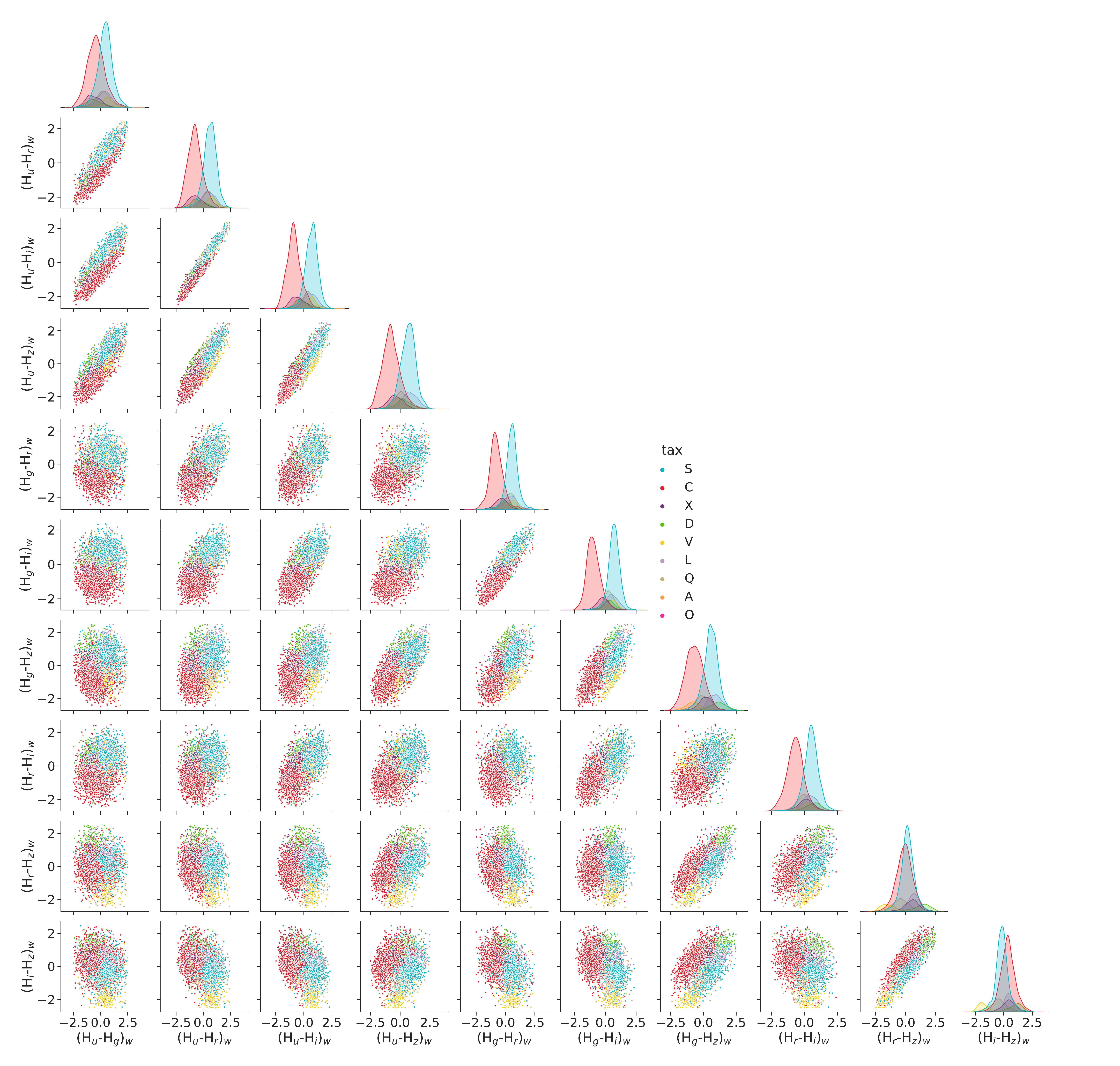}
    \caption{Correlation between different absolute colors of Sloan. The color scheme corresponds to C10's classification.}
    \label{parameters}
    \end{figure}

\end{appendix}
\end{document}